\documentclass[12pt]{article}
\usepackage{amssymb, amsthm, amsmath}

\renewcommand{\r}{\mathbf{r}}
\newcommand{\h}{\mathbf{h}}

\def\g{\ifmmode{g \hskip -7pt g}
     \else{\hbox{$g \hskip -7pt g$}}\fi}         
\newcommand{\Radon}{{\mathcal{R}}}
\newcommand{\XRay}{P}
\newcommand{\s}{\mathbb{S}}
\newcommand{\Zs}{\s ^2}
\renewcommand{\H}{\s ^2_{\h}}

\renewcommand{\phi}{\varphi}
\renewcommand{\epsilon}{\varepsilon}

\begin{document}
\title{A one-dimensional Radon transform on $\mathbf{SO}(3)$ and its application to texture goniometry}
\author{Swanhild Bernstein \thanks{Bauhaus University Weimar, Faculty of Media, Bauhausstr. 11,
D-99423 Weimar, Germany, email: {\tt swanhild.bernstein@fossi.uni-weimar.de,} phone: +49 3643 58 3716, fax: +49 3643 58
3709, and University of Mining and Technology, Fak. 3, Mathematische Geologie und Geoinformatik, D-09596 Freiberg,
Germany,} \and  Helmut Schaeben
\thanks{University of Mining and Technology, Fak. 3, Mathematische Geologie und Geoinformatik, D-09596 Freiberg,
Germany, email: {\tt helmut.schaeben@geo.tu-freiberg.de}} }
\date{~}
\maketitle The full paper will be published in \\[1ex]
  \centerline{Mathematical Methods in the Applied Sciences.}
\section{Introduction}
\subsection{Motivation from texture goniometry}
Texture analysis with X-ray diffraction data is the analysis of the orientation distribution by volume and asks for a
measure of the volume portion $\Delta V/V$ of a polycrystalline specimen of total volume $V$ carrying crystal grains
with orientations within a range (volume element) $\Delta G\subset G$ of the subgroup $G$ of all feasible orientations
$G\subset \mathbf{SO}(3).$\\[1ex]
The orientation $\g$ of an individual crystal in a polycrystalline specimen is the active rotation $\g \in
\mathbf{SO}(3): \, K_{\cal S} \mapsto K_{\cal C}$ that maps a right--handed orthonormal coordinate system $K_{\cal S}$
fixed to the specimen onto another right--handed orthonormal coordinate system $K_{\cal C}$ fixed to the crystal,
\begin{equation} \label{defg}
\g \; K_{\cal S} = K_{\cal C}, \enspace {\g} \in \mathbf{SO}(3) \,.
\end{equation}
If a unique direction is represented by unit vector ${\h}$ with respect to the crystal frame $K_{\cal C}$, and by unit
vector ${\r}$ with respect to the specimen frame $K_{\cal S}$, then the coordinates of the unique direction transform
according to
\begin{equation} \label{transformg}
{\r}_{K_{\cal S}} = \g \; {\h}_{K_{\cal C}} \,.
\end{equation}

The commonly applied convention in texture analysis (H.J. Bunge, \cite{bu1}; \cite{bu2}) refers to the notion of
passive rotation and Eq.~ \ref{transformg} is written in the form
\begin{equation} \label{transformbunGe}
{\h} = g {\r}
\end{equation}
where obviously $g = {\g}^{-1}$. Since we aim at a unified view of inversion formulae developed in such apparently
diverse filds as texture analysis, integral geometry, and spherical tomography, we use here the notation
Eq.~\ref{transformbunGe} familiar in applied sciences. Thus, it is our hope to accomplish clarification without
confusion by yet another convention.\\[1ex]
Assuming that the measure possesses a probability density function $f: G\mapsto \mathbb{R}^1_+ ,$ then
$$ prob(g\in\Delta G)=\int_{\Delta G} f(g)\,d\omega _g $$
and $f$ is referred to as the \emph{orientation density function} by volume and $d\omega_g =
\sin\beta\,d\alpha\,d\beta\,d\gamma $ is the usual Riemannian measure of $\s^3$ which differs from the invariant Haar
measure $dg$ of $\mathbf{SO}(3)$ by a constant factor, we have $d\omega _g=8\pi ^2 dg.$\\[1ex]
In X-ray diffraction experiments the orientation density function $f$ cannot be directly measured but with a texture
goniometer only pole density function $\XRay (\h,\,\r)$ can be sampled, which represents the probability that a (fixed)
crystal direction $\h $ or its antipodal $-\h $ statistically coincide with the specimen direction $\r .$ With respect
to the experiment the feasible crystal directions are the normals of the crystallographic lattice planes. A \emph{pole
density function} is the tomographic projection of an orientation density function which is basically provided by
$$ \XRay f(\h,\,\r)=\frac{1}{2} \left( \Radon (\h,\,\r)+\Radon (-\h,\,\r)\right) $$
with
\begin{align*}
\Radon f(\h,\,\r)&=\frac{1}{2\pi }\int_{\{g\in
\mathbf{SO}(3):\,\h=g\,\r\}} f(g)\,d\omega_g  \\
&=4\pi \int_{SO(3)}f(g)\delta_{\r}(g^{-1}\cdot \h )\,dg = (f\ast \delta_{\r})(\h ). \nonumber
\end{align*}
\section{Main results} We start with the  one-dimensional Radon transform on $\mathbf{SO}(3)$ and its
inversion. Using a group-theoretical approach we will obtain an inversion formula which will be the basis for other
inversion formulae. An important tool will be the series expansionsin surface harmonics on $\s ^2\times \s ^2$ as well
as in rotational
harmonics on $\mathbf{SO}(3)$.\\[2ex]
Let $f\in L^2(\mathbf{SO}(3)).$ We define the \emph{Fourier series} on $\mathbf{SO}(3)$ by
\begin{align*}
f(g) &= \sum_{l=0}^{\infty} \sum_{m=-l}^l\sum_{n=-l}^l (2l+1)\,\hat{f}_l^{mn} \,\overline{(D^l_{m,n})(g)}=
\sum_{l=0}^{\infty} \sum_{m=-l}^l\sum_{n=-l}^l (2l+1)\,\hat{f}_l^{mn} \,D^l_{m,n}(g^{-1}),
\end{align*}
where
$$ \hat{f}^l_{mn} =\int_{SO(3)} f(g)\,D_{m,n}^l(g)\, dg,\ -l\leq m,n \leq l,\ l=0,1,\ldots $$
and $dg$ is the Haar measure on $\mathbf{SO}(3).$\\[1ex] Therefore we get
$$\mathcal{F}_{SO(3)}f = 4\pi\,\mathcal{F}_{\s ^2\times \s ^2}(\Radon f),$$
i.e. the Fourier coefficients of $f$ in $L^2(\mathbf{SO}(3))$ are equal to $4\pi$-times the Fourier coefficients of
$\Radon f$ in $L^2(\s ^2\times \s ^2).$ Furthermore, we have several inversion formulae
\begin{align}
f(g) &=4\pi \check{\Radon}(-2\Delta_{\s^2\times\s^2}+1)^{1/2}\Radon f \label{a}\\[1ex]
     &=4\pi (-4\Delta _{SO(3)}+1)^{1/2} \check{\Radon}\Radon f \label{bp}\\[1ex]
     = \frac{1}{4\pi}&\left\{ \int_{\Zs} \Radon (\h,\,-g\h )\,d\H + 2\int_0^{\pi} \cos \frac{\theta }{2} \,\frac{d}{d \cos
\theta } \int_{\Zs} (Wf)(\h,\,g\h,\,\cos \theta )\,d\H \,d\theta \right\}, \label{sm}
\end{align}
where $\check{\Radon} $ denotes the dual Radon transform in $L^2.$\\[1ex]
Formula (\ref{a}) is the starting point of all other inversion formulae which are derived by series expansion into
spherical harmonics and rotational spherical harmonics. The second formula (\ref{bp}) is the analog of the so-called
backprojection formula in the Euclidean case and could be used to detect edges and wedges. The third formula is of
special interest because it occurs in crystallographic literature  see for example \cite{bu1}, \cite{bu2},
\cite{sm}.\\[1ex]
Further we demonstrate its equivalence to the inversion formula of the spherical Radon transform on $\s ^3$ given by
S.~Helgason (\cite{he1}, \cite{he2}).\\[1ex]
The X-ray transform in texture goniometry differs from the Radon transform on $\mathbf{SO}(3)$ in the following way.
Due to Friedel's law which states that the X-ray cannot distinguish between the top and the bottom of the lattice
planes, we are only able to measure a mean value which correspondence to a negligence of the orientation on
$\mathbf{SO}(3)$. Therefore the X-ray transform in texture goniometry is not an isomorphism.

\thebibliography{999}
\bibitem{bu1} Bunge HJ. {\it Mathematische Methoden der Texturanalyse,} Akademie-Verlag Berlin, 1969,
\bibitem{bu2} Bunge HJ. {\it Texture Analysis in Materials Science,} Mathematical Methods, Butterworths,
              London, Boston, 1982,
\bibitem{he1} Helgason S. {\it Groups and Geometric Analysis,} Integral Geometry, Invariant Differential Operators,
and Spherical Functions, Academic Press, San Diego, New York, Boston, 1984,
\bibitem{he2} Helgason S. {\it The Radon Transform,} Second Edition, Birkh\"{a}user, 1999,
\bibitem{sm} Matthies S. {Aktuelle Probleme der quantitativen Texturanalyse,} ZfK--480, Zentralinstitut f\"{u}r
Kernforschung Rossendorf bei Dresden, ISSN 0138--2950, August 1982.
\end{document}